# REVIEWS

# Water Movement in Vascular Plants: A Primer

Sanjay P. Sane* and Amit K. Singh

Abstract | The origin of land plants was one of the most important events in evolutionary history of earth in terms of its broad impact on metazoan life and the biotic environment. Because vascular tissues enabled land plants to meet the challenges of terrestrial life, it is important to understand the mechanistic basis of water transport through these tissues from soil to the canopy of trees, in some cases almost 100 meters high. The answers to these questions involve not only the biology of plant vasculature, but also the physical properties of water that enable such transport. Although early researchers proposed the hypothesis of cohesion-tension of water as the likely mechanism for sap ascent, the exact mechanism of transport continues to be a hotly debated topic in the field of plant physiology. This debate continues to be enriched with several sophisticated studies on plants of various morphologies growing in diverse habitats. Although a wealth of evidence has upheld the cohesion-tension theory as being fundamentally correct, several important details about how plants deal with vascular stress remain unknown. Here, we review the recent literature on this topic with the aim of highlighting how a multi-disciplinary perspective has contributed to our understanding of the cohesion-tension theory of sap ascent in plants.

## 1. Introduction

The invasion of land by plants between 480 to 360 million years ago (Kenrick and Crane, 1997) is arguably one of the most significant events in the biological history of earth. The algal precursors of modern land plants were perennially submerged in water, thus enabling all their tissues to be hydrated according to their need. However, these organisms faced several limitations to the expansion of their range and size within their submerged environs. Contrary to common perception, water is not entirely colorless but absorbs some red-green radiation that is required for normal photosynthetic activity (Niklas, 1992). Thus, there is a decline in the intensity of solar radiation with increasing depth of water, which restricted the algae to only the surface layers of water. The lack of a water transport system within the algal tissues also meant that they were severely constrained in size above water, as also were the early land plants such as liverworts or mosses which had a rudimentary capillary-like water transport system (Sperry, 2003).

The land invasion of plants mitigated many of these problems, but in turn presented new challenges such as robust anchorage, defense from dessication, avoiding excessive cooling due to transpiration etc. For these reasons, the evolution of a vascular transport system in plants was perhaps the key step in the successful land invasion of plants. Yet, although this topic is fundamental to our understanding of basic functioning of land plants, the mechanism of vascular transport remains one of the most debated topics in the field of plant physiology (for reviews see Sperry, 2003;

National Centre for Biological Sciences, Tata Insititute of Fundamental Research, GKVK Campus, Bangalore 560065, India
*sane@ncbs.res.in

Xylem: Xylem is one of the complex vascular tissue types in higher plants. Its primary function is to transport sap from roots to stem and leaves, thereby replacing water lost during transpiration and supplying minerals. In addition, it also provides mechanical support. Typically, xylem is made up of dead cells that make up vessels and/or tracheids with parenchyma cells and wood fibers.

Osmotic potential: Osmotic potential or Osmotic pressure is the external pressure that needs to be applied to a solution to stop water influx from another solution separated by a semi-permeable membrane.

Van der Waal's forces: van der Waals forces (discovered by Dutch scientist J. D. van der Waals) result from attractive and repulsive electromagnetic interactions between or within molecules. They are distinct from covalent or electrostatic ionic interactions and are composed of forces between permanent dipoles (Keesom forces), or permanent dipoles with their induced dipole (Debye forces), or two transiently induced dipoles (London force).

Steudle, 2001; Tyree and Zimmermann, 1983). This debate has continued at many levels - from the basic physical properties of water and nature of its motion in capillaries, to which experimental techniques are most appropriate for measuring internal pressures in the vascular structure of plant xylem, and what is the biological response of plants to stresses due to water shortage or excess salinity. In this article, we review the history and recent literature on water transport in plants with a focus on the tools and techniques and major experimental challenges in the field.

### 1.1. The Physical Properties of Water

The ubiquity of water often makes us lose sight of the fact that the physical properties of water are rather anomalous as compared to other liquids of its kind. Water itself is odd because it is liquid at room temperature, in sharp contrast to its neighboring elements in periodic chart such as carbon, nitrogen, fluorine etc. that form gaseous compounds with hydrogen. Because water molecules are polar in nature, they form strong hydrogen bonds with strengths of about 23.3 kJ/mol, which are 2 - 5 times stronger than the intermolecular Van der waal's forces (Niklas, 1992). Perhaps as a direct consequence of such molecular interactions, water can withstand a high tensile stress as compared to other liquids of its kind, a property of critical importance for vascular transport of water.

The first measurements of tensile strength of water had several surprises to offer. Following in the footsteps of some pioneering experiments by Osborne Reynolds on internal cohesion of liquids (Reynolds, 1877), Lyman Briggs developed a Z-shaped capillary that could be spun around its axis, thus exerting high tensile forces at the center of the column (Briggs, 1950; see also a critical evaluation of various methods of measurements of internal tensions by Temperley, 1946; Temperley, 1947; Temperley and Chambers, 1946). Using this method, Briggs measured the tensile strength of water and showed that it could withstand internal tensions of nearly -26 MPa at room temperatures before it ruptures or cavitates (Smith, 1991). Subsequent measurements of the internal tensions in water have yielded values as high as -140 MPa (Zheng et al., 1991). These measurements are of direct relevance to the vascular transport of water in plants because they directly address how high a column of water can be pulled up using negative pressures, without loss of integrity.

### 1.2. Water Potential

Movement of water within plants is primarily dictated by a physical quantity called water potential ($\psi_w$), defined as the potential energy per unit volume of water. Water always flows from regions of high water potential to regions of low water potential. The net water potential of a system is typically expressed as a sum of several components:

$$\psi_w = \psi_s + \psi_m + \psi_p + \psi_g \qquad (1)$$

where $\psi_s$ is the potential energy contribution due to solute concentration (for very dilute solutions, the same as osmotic potential $cRT$, where $c$ is the solute concentration, $R$ the gas constant and $T$ the temperature in Kelvin), $\psi_m$ the contribution due to the cohesive and tensile interactions of water with the matrix within which it is transported, $\psi_p$ the pressure potential due to external pressure differences, and $\psi_g$ the gravitational potential energy (equal to $\rho gh$), where $\rho$ is the density of water, $g$ the acceleration due to gravitation and $h$ is the height of the water column. Because there can be no movement of water into pure water under normal conditions of pressure and temperature, the water potential for pure water is assigned the arbitrary value of 0. Thus, in any scenario (solutions, liquid within any matrix, or under gravity etc.), water moves from less negative to more negative water potential (Niklas, 1992; Tyree, 1997).

#### 1.2.1. Cohesion-Tension in Xylem Vessels

Several years before Briggs' measurement of internal tension in water, Dixon and Joly (1900) and Askenasy (1895) had independently proposed that water inside the xylem vessels is transported under a state of high tension because it is pulled up rather than pushed up as a result of transpiration. Dixon and Joly (1900) likened it to the ascent of liquid in a porous vessel, which contains small sized pores and hence are able to collectively withstand pressures of large magnitudes. They argued that the loss of water through transpiration from stomata was similar to the porous vessel situation and hence set up an internal tension that enabled water to be pulled up through the xylem as it evaporated from the leaves during transpiration. For this hypothesis to hold true, the xylem must be able to exert negative pressures of large magnitudes to be able to pull a water column through tens of meters—difficult even for a vacuum suction pump (see for example, Hayward, 1970). Even discounting the contributions of $\psi_s$, $\psi_m$ and $\psi_p$ in equation (1), a negative pressure of at least -0.1 MPa (or –1 atmosphere) is required to pull up a column of liquid up to the canopy of a 10 meter tree. In fact, because $\psi_s, \psi_m$ and $\psi_p$ are much greater in magnitude than the gravitational

**Halophytic plants:** Plants such as the mangrove plants that can survive in conditions of extreme salinity.

**Phloem:** Unlike xylem which is largely composed of dead tissue, phloem is a living vascular tissue that transports organic nutrients (e.g. glucose) to various parts of the tree.

**Angiosperms:** Angiosperms are flowering vascular plants or trees, such as all fruit trees, in which the seeds are covered.

**Pteridophytes:** Pteridophytes are vascular plants such as ferns, which have no seeds, but reproduce through spores.

**Gymnosperms:** Gymnosperms are non-flowering vascular plants or trees such as conifers and cycads, in which the seeds are not covered.

potentials, the actual internal tensions required to pull up a water column are much higher. In extreme conditions, such as those experienced by halophytic plants within mangroves which grow in hyper-osmotic conditions, the required internal tensions may be as high as a -1 to -2 MPa for proper water transport. Likewise, the internal tensions required for very tall trees, such as the California Redwoods which can grow up to 100 m may be estimated to be on the order of a -2 MPa (Scholander et al., 1965).

It is important to emphasize here that within an intact xylem tissue, water is thought to experience negative *i.e.* below vacuum pressures. This concept should not be confused with suction forces generated by standard vacuum pumps, which at best, can yield near zero but positive values of pressure. In the latter case, a pressure difference forces the liquid to move towards the lower pressure region, but it does not hold the liquid itself under any significant tension. However, a column of liquid held under tension can pull back on an external extensive force (Herbert and Caupin, 2005). Similarly, water being pulled up due to transpiration in plants is under tension and hence in a metastable state *i.e.* it often experiences pressures under vapor pressure without evaporating.

Thus, for the cohesion-tension theory to explain the transport of water within plants, at least three conditions must be satisfied. First, cohesive forces of water should be high along with high tensile strength. Second, there must be hydraulic continuity between the roots, stem and the leaves at all times. Third, there must be a sufficient gradient of water potential between the leaves and the roots to pull up water through the required height of the column. As evident from Briggs' experiments, a pure and degassed water column can withstand tensions of up to -26 MPa. Under such high values of tension, water with dissolved gases has a tendency to evaporate thus forming bubbles within the water column that can disrupt hydraulic continuity. Briggs' experiment showed that the water's molecular structure imparts it with enough strength to withstand the tensions required to pull a water column up to the height of the tallest trees. Under such conditions, it can also exert an inward pull on the wall of the xylem tube, which requires natural reinforcement in the form of thickened internal walls to withstand such inward forces (Tyree and Zimmermann, 1983). Hence, in addition to the above intrinsic factors, xylem tracheid morphologies are also important for proper ascent of sap (e.g. Steudle, 2001) as described in the following section.

### 1.2.2. *Morphology of the Vascular Structure of Xylem*

Vascular tissues are primarily composed of *xylem* which transports sap and mineral nutrients from roots to the canopy, and *phloem* which transports the photosynthetic products such as sugar from the leaves to the rest of the tree. Xylem is composed of a network of sap-conducting micro-capillaries which vary in length and organization, in addition to non-conducting elements like parenchyma cells and fibers (Fig. 1A). If water enters the root hair through the plasma membrane into the cell cytoplasm and then goes from cell-to-cell *via* connecting pores (called *plasmodesmata*), the transport is called *symplastic*. Alternatively, if water does not enter the plasma membrane and passes instead through extracellular spaces, the transport is called *apoplastic* (Fig. 1B).

Within the stem xylem, the cells that make up the individual tubes, called *tracheids*, are devoid of any protoplasm and hence essentially dead. However, their thick lignified cell walls survive to form elongated tubes tapered at each end. They have a primary cell wall which lacks perforations. The secondary cell walls differ greatly in morphology leading to diverse tracheid architectures. Retracing the fossil ancestors of land plants reveals that the tracheid architecture in early land plants showed remarkable diversity of shapes ranging from simple capillary-like tubes in bryophytes, the S-type cell wall with distributed micropores derived from plasmodesmata and interspersed with spongy thickening of the walls, the G-type cells walls with regular thickening similar to modern plants, and the ladder-like pitted P-type tracheids (see Fig. 5 in Kenrick and Crane, 1997; Sperry, 2003). The tracheid elements are lined on the inside by secondary cell walls with annular, spiral, reticulate, or pitted wall thickenings among other variants (Fig. 1C). The pits form numerous perforations which enable lateral transport of water between adjoining tracheids (Frey-Wyssling, 1976). Sap flows through these pits more efficiently in longer vessels than in shorter ones because transport of sap through bordered pit pairs causes a greater resistance in water flow (Zimmermann and Potter, 1982). This organization enables a continuum of vessel elements divided at regular intervals by perforation plates of different geometry and porosity.

Unlike tracheids which are found in almost all vascular plants, vessel elements occur in most angiosperms but only a few pteridophytes or gymnosperms. Vessel elements contain perforated end walls and the middle lamella and can thus be stacked lengthwise to form long vessels. In absence of the intervening pit membrane, longitudinal water flow is more unimpeded through vessels than through tracheids. Lateral water flow, which is mediated by tracheids, is however not mechanistically very different in angiosperms *vs.* gymnosperms (Niklas, 1992). Among flowering plants, the vessels in monocotyledonous plants tend to be scattered, whereas vessels in dicotyledonous stems are typically arranged in neat circular bundles. Even within the same species and same plant, vessel conduit architecture may vary from diameters as lows as 2–3 microns to as high as 100s of microns depending on size and location.

Figure 1: Morphology related to water transport in plants. **A.** Cross-section of a typical wood stem. The parts relevant to water transport are labeled. **B.** Water transport pathways through root hair. Red lines show both the symplastic and apoplastic routes of water transport through roots. **C.** Diversity of tracheid types. The morphology of the internal walls of tracheids influences the transport of water through the tracheid. **D.** Brodersen et al's (2010) proposed mechanism for embolism repair. According to this model, embolisms cause neighboring cells to release solutes into the affected vessel. The resulting osmotic gradient causes water movement from surrounding fibers and cells into the vessel. Eventually, the water droplets span the whole width of the vessel. Because the water in all vessels, including neighboring ones, is under a state of negative pressure, these water droplets may move to neighboring vessels, keeping the embolism in place. If however, there is net water accumulation within the embolised vessel, and the gas present in that vessel will eventually dissolve to repair the embolism (Brodersen et al., 2010).

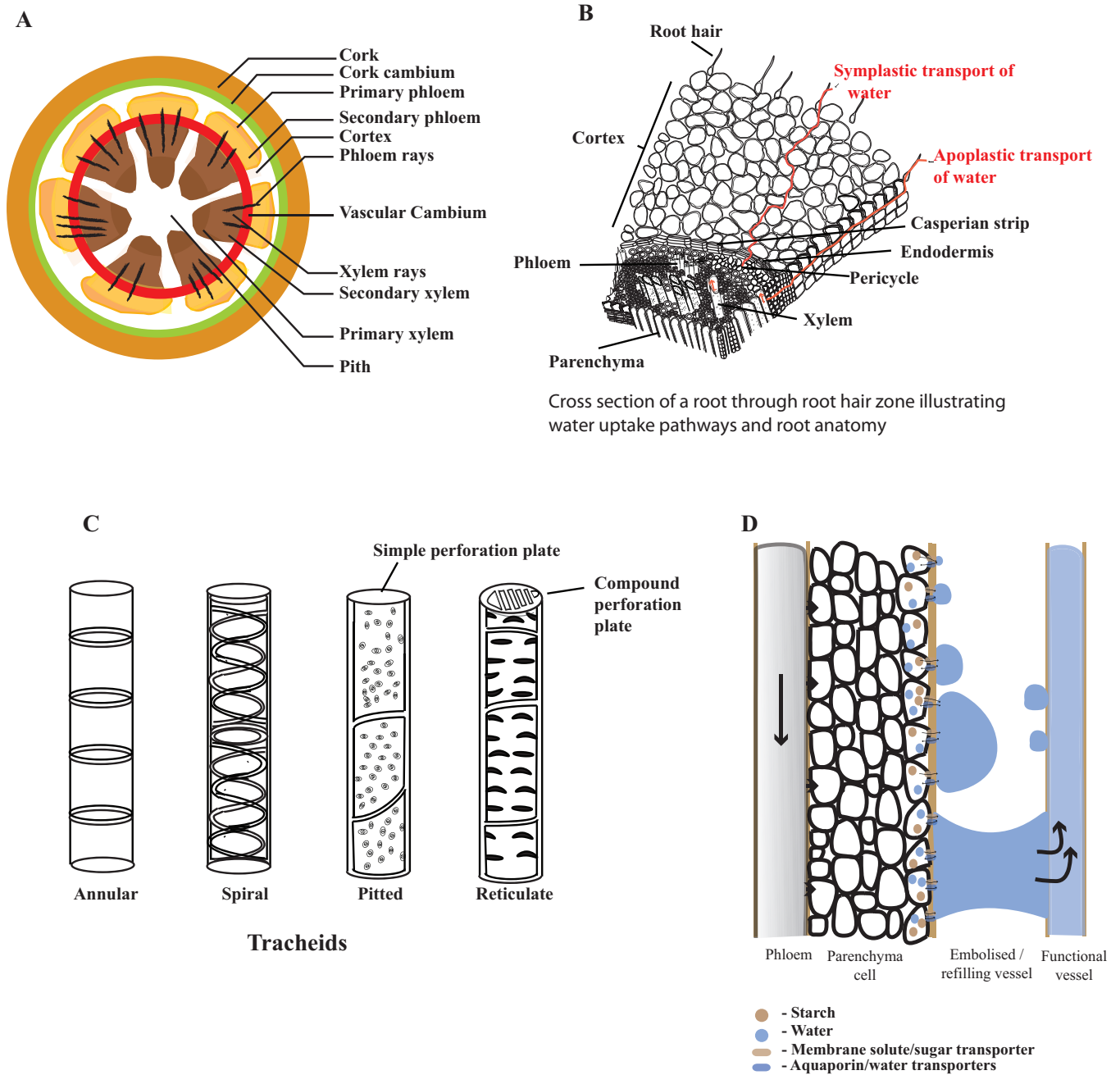

The earliest observations of such spatial patterns within tracheid architecture were described by Karl Sanio for gymnosperms and later termed as Sanio's laws. Sanio described five such rules. First, tracheids increase in size from the center of the tree to the periphery with successive annual growth rings until, beyond a certain size, their size plateaus. Second, the constant size in the outermost rings increases up to a certain height and there after diminishes. Third, the tracheids in branches are smaller in size but correlated to those in stem at that level. Fourth, the tracheids of plateau size in the knotted branches at the top of a tree increases and decreases towards the periphery, but are more irregular than in the stem. Fifth, the diameter of the vessels in roots also increases and decreases and then plateaus (Sanio, 1872). However, more recent evidence on such patterns suggests that such organizing principles of hydraulic architecture must be treated as thumb rules rather than laws (Mencuccini et al., 2007). Not surprisingly, the complexity of vessel and tracheid morphologies substantially hinders the theoretical treatment of ascent of sap.

### 1.3. Basic Principles of Hydraulic Conductance

Because of the capillary-like nature of the conduits, flow in the xylem is often approximated as capillary. Based on this assumption, the flow rate through the capillary ($dV/dt$) is directly proportional to the applied pressure gradient ($dP/dl$) across the length of the capillary $l$. Thus,

$$\frac{dV}{dt} = L_p \frac{dP}{dl} \qquad (2)$$

where $L_p$, the net hydraulic conductance is often approximated to the standard Hagen-Poiseulle equation

$$L_p = \frac{\pi R^4}{8\mu}, \qquad (3)$$

where $R$ is the radius of the tracheid vessel and $\mu$ is the dynamic viscosity. As is clear from equations 2 and 3, the flow rate through a tube is directly proportional to the fourth power of the vessel radius leading to the theoretical expectation that in order to maximize flow, plants should increase the diameter of their vessels. This should especially be the case for tall trees where maximizing flow rate may be critical.

However, detailed comparative surveys of vessel morphology suggest that plants appear to trade-off vessel diameter with vessel number. In many cases, we see a larger number of small-diameter vessels rather than a fewer large-diameter vessels. Tyree and Zimmermann (1983) suggest that this may be because vessel number also trades-off with safety for the plant. Because larger diameter vessels are more likely to cavitate due to collection of gas from a larger pool of bubble seeds, their cavitation would substantially impact the hydraulic continuity leading a catastrophic loss of water supply. In contrast, the presence of many small-diameter vessels means that the loss of a few due to cavitation or parasitism may impact hydraulic continuity of a few vessels at a time, but may not greatly influence water transport in the whole tree (Tyree and Zimmermann, 1983).

Due to the complexity and diversity of internal structures of tracheids and because the flow characters are influenced by internal surface properties of the tubes, equation 3 is, at best, only a loose approximation of the hydraulic conductance through xylem. Further complexity is added by the fact that xylem is not a system of passive tubes that conduct water, but contain hydrogels such as pectins, that can actively swell and deswell with pH changes in the sap, thereby modulating changes in hydraulic resistance (Zwieniecki et al., 2001). In absence of a proper theoretical framework, our understanding of the conductance of sap through xylem rests largely on several detailed experimental measurements and characterization of the internal pressures within the xylem.

### 1.4. Embolism and its Repair

In the above sections, the water was usually assumed to be de-ionized and containing no dissolved gases. Under such conditions, a water column can attain higher tensions without the risk of cavitation or nucleation. However, neither of these conditions is obeyed in nature with the consequence that the value of internal tensions that a water column can withstand is significantly less than the empirical values suggest. When the internal tension exceeds the value that can be withstood by water, dissolved gases begin to accumulate at the solid-liquid interface in the form of tiny bubbles which eventually coalesce and cause the hydraulic continuity to break down.

*The Hagen-Poiseuille:* The Hagen-Poiseuille equation describes the pressure drop in a fluid flowing through a long, regular cylindrical pipe whose length far exceeds its diameter. It assumes that the flow is incompressible, laminar and viscous.

This condition is called *embolism*. Because hydraulic continuity is essential for nutrient transport, any breakdown in its continuity means that the part of the plant or tree above the embolism is deprived of nutrition unless hydraulic continuity can be restored. Other causes of embolism include extreme osmotic stress, drought, or freeze-thaw events during winter. In many cases, plants have evolved mechanisms to remove embolism and restore water transport in the cavitated vessels, including the ability to refill embolised conduits while maintaining the tension in the sap in the xylem. (Salleo and Gullo, 1986; Sperry et al., 1987; Tibbetts and Ewers, 2000; Holbrook et al., 2001; Clearwater and Goldstein, 2005).

How plants detect an embolism and whether or not its repair is an active biologically driven process remain open questions (Salleo et al., 2007; Tyree and Sperry, 1989; Tyree et al., 1999). To understand how embolism repair and refilling of vessels under tension works, it is necessary to understand the origin of the driving force that impels water into cavitated conduits, the source and pathway of water for refilling, and the physical nature of the hydraulic compartmentalization which separates refilling and functional conduits. Using state-of-art Magnetic Resonance Imaging (MRI), Holbrook and colleagues (2001) observed the *in vivo* repair of cavitated vessels and the subsequent restoration of their leaf potentials (also see Scheenen et al., 2007). More recently, Bordersen and colleagues (Brodersen et al., 2010) have performed 3D high resolution computer tomography (HRCT) on intact grape vines to demonstrate the active repair of embolized vessels. They showed that individual droplets of water in the embolized vessels expanded to eventually fuse and force the redissolution of trapped gases. Embolism repair may also be mediated by root pressure and through active participation of cells present around the embolised conduit (Figure 1D; Salleo et al., 1992; Salleo et al., 1996; Holbrook and Zwieniecki, 1999; Zwieniecki and Holbrook, 2009; Brodersen et al., 2010).

### 1.5. Measurement of the Xylem Pressures

Two main techniques have been used to measure internal xylem pressures:

*Pressure Chamber Measurements*
The first technique, modified by Scholander (Scholander et al., 1965) from an earlier version used by Dixon and Joly (1895), works on the assumption that when a plant part such as a leaf or a stem or twig is rapidly excised from a plant, the internal tension in the xylem pulls the water from the xylem tracheids into the cells until they are in equilibrium. If such a leaf is now inserted within a closed pressure chamber with only its cut part exposed to ambient atmospheric pressure and the surface of the leaf is subjected to known external pressures, the external pressure potential ($\psi_p$) is such that water moves out of the cells and back into the xylem until its meniscus is visible at the cut surface. The externally applied pressure at this point may then be said to be equal and opposite in magnitude to the tension within the xylem. Because the excised leaf transpires and can very quickly build high internal tensions, it is necessary to conduct these experiments in a short time window immediately following excision and then ensuring that the surroundings are sufficiently humid to prevent water loss.

Scholander and colleagues (Scholander et al., 1965; Fig. 2A) used this technique extensively to survey the negative internal pressures of various plants, including some called epiphytes, which are parasitic on other plants. As expected the parasitic epiphytes have a significantly more negative internal pressure as compared to normal plants. The range of values obtained from such measurements (-0.5 to -8 MPa) also appeared to be in general agreement with the pressure values required for sap ascent. Not surprisingly, this technique quickly gained rapid currency and was adopted by several plant physiologists to measure and compare the water potentials of plants in diverse range of habitats ranging from deserts and mangroves *vs.* more moderate climes, plains *vs.* hills and mountains, summer *vs.* freezing winters, as well as plants of diverse heights from small saplings to tall conifers (Scholander et al., 1965). The typical measurements from such studies offered values that made sense. For instance, when comparing parasitic plants such as mistletoe with their hosts, Scholander et al. (1965) showed that suction of the parasitic plants exceeds their hosts by nearly 0.1 to 0.2 MPa. Similarly, the range of measurements of water potential in mangroves was −3.5 to −6 MPa, much more negative than desert, forest or fresh water dwelling plants.

The pressure chamber technique is obviously invasive because it involves severing the leaf or twig for the experiment. It has also been criticized for being error-prone because the pressure seal that isolates the part of the plant inside the chamber from the outside also exerts lateral pressure on the twig which can alter the pressure values required to recover the water meniscus at the cut end.

*Pressure Microprobe Measurements*
The pressure-chamber based results described above were challenged by Balling and Zimmermann (Balling and Zimmermann, 1990; Niklas, 1992),

**Epiphytes:** Plants or creepers, such as ivy, that grow on other plants. They usually derive their nutrition independently of their host plant, and are not parasitic.

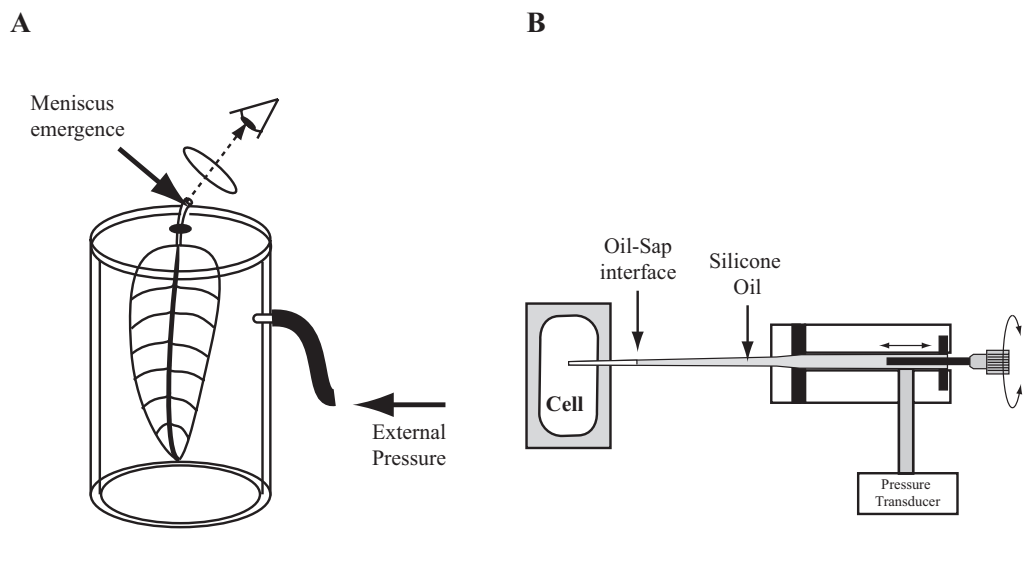

Figure 2: Techniques of pressure measurements in plants. **A.** *Scholander pressure chamber method:* Originally based on a previous method used by Dixon and Joly (1900), the Scholander pressure chamber method was the first to successfully demonstrate and measure negative sap pressures in plants. However, they are invasive and error-prone due to the requirement that the plant leaf or twig be detached from the plant for the purpose of the experiment. **B.** *Pressure microprobe technique:* A fairly non-invasive technique to measure the internal pressures in plant xylem without significantly altering its structure. It is sensitive at considerably lower range of pressures than the Scholander technique. It is also relatively more difficult to position within a plant tissue.

who used a less invasive method to determine sap pressures in xylem of *Nicotiana* plants. Their method used a microprobe, similar to those typically used to determine turgor pressure of cells (Green and Stanton, 1967) or the internal pressure of modified leaf bladders in *Utricularia* (Singh et al., 2011).

In this method, a microcapillary tip is filled with water and a marker of some sort, either a trapped air bubble or an oil-water interface, to monitor changes in volume or pressure. In the case of a trapped air bubble, we can monitor the changes in the bubble volume due to interaction with the external pressures. If the external pressure is lower (higher) than the internal pressure of the bubble, the bubble expands (contracts) and the change in volume can be monitored using a microscope. The external pressure can then be calculated using the standard Boyle's law $P_0 V_0 = P_1 V_1$, where $P_0$ and $V_0$ are the initial values of volume and pressure and the $P_1$ and $V_1$ are the final values (see supplementary information, Singh et al., 2011).

When the volume of cell is small, even the few microlitres of water entering the microcapillary during measurement can alter the pressures within the cell, causing large errors. To keep careful track of the changes in volume, such micro-capillaries are typically filled with oil to keep careful track of the oil-water interface, which is unfortunately also a likely zone for bubble formation during measurement of high negative pressures. When using microprobes to measure the states of systems with larger volumes (e.g. internal pressures of xylem), the small changes in volume are negligible. In such cases, the measurement can be carried out without fear of changing the overall state of the system. However, one disadvantage of this method is that it cannot measure instantaneous changes in pressure due to the time taken by the capillary system to equilibrate. However, quasi-steady measurements are easily possible using this method. The microcapillaries used in modern pressure probes are additionally fitted to sensitive pressure transducers to enable measurement and amplification of the smallest changes in state of the measured system (Tomos and Leigh, 1999).

In addition to the above methods, there are also thermal methods of determining sap flow through trees. One method is to estimate sap flow by measuring how moving sap influences the heat balance of a heated stem section. Alternatively, the researcher may deliver a heat pulse into the stem and measure how the heat pulse is conducted by the moving sap using thermocouples that are placed at a specific distance $x$, upstream and downstream to the site of the heater. The rate of flow $v_h$ can then be calculated by the equation

$$v_h = \frac{k}{x} \ln\left(\frac{\delta T_1}{\delta T_2}\right) \qquad (4)$$

where, k is the thermal diffusivity coefficient and $\delta T_1$ and $\delta T_2$ are the temperature measurements upstream and downstream of the heater (Clearwater et al., 2009).

Because these methods involve insertion of the both the heaters and thermocouples into the tree sapwood, they are particularly useful in cases where the stem is somewhat thick (Swanson 1994; Smith & Allen 1996).

**1.6. The Controversy Over Ascent of Sap Measurements**

In the first, direct *in vivo* measurements of the xylem pressures using the pressure microprobe, Balling and Zimmermann (1990) filled the microprobe with degassed water to avoid any cavitation or bubble-formation at the oil-water interface. Using this device, they measured the internal pressures of *Nicotiana* plants and compared their measurements with values obtained from the Scholander pressure chamber. These comparisons revealed that although both sets of measurements were internally consistent, the two methods showed very different absolute values of internal pressures. Thus, although the pressure probe measurements confirmed the tension in xylem sap, it recorded values that were less negative by nearly an order of magnitude or even positive (between 0 and 0.1 MPa) as compared to the Scholander pressure chamber measurement, which measured absolute values on the order of −0.1 MPa. They used several experimental controls to double-check that their pressure probe measurements were accurate.

The pressure probe measurements in *Nicotiana* and other plants measured over the years challenge the existing set of ideas about ascent of sap (Balling and Zimmermann, 1990; Tomos and Leigh, 1999). First, the xylem tension seemed unaffected by transpiration, thus calling into question the Cohesion-Tension theory of ascent of sap. Second, when simultaneous measurements of xylem pressure were carried out by placing the whole plant along with the microprobe inside the Scholander pressure chamber, they did not find a one-to-one correspondence between pressurization of the chamber and the change in microprobe measurements. Indeed, the microprobe appeared to be insensitive to even steep increases in external pressures on the whole plant. However, if the above experiments were repeated with roots cut and the stem inserted directly into water, a one-to-one relationship could be observed between pressure probe measurements and actual pressure. This meant that the insensitivity of the microprobe to the external changes in pressure on an intact plant was not an artifact, but rather that external pressure was not directly transmitted to the xylem.

This finding negated the fundamental premise upon which the pressure chamber experiments were based. Later measurements in sugarcane and maize plants indicated, however, that the one-to-one relationship between xylem pressure probe measurements and Scholander pressure chamber measurements does hold for non-transpiring, but significantly deviates for transpiring leaves (Melcher et al., 1998). Together, these results presented a somewhat conflicted picture about the mechanisms underlying ascent of sap in plants and researchers using pressure probes argued that cohesion tension theory may not be the primary mechanism for the ascent of sap.

Some resolution to this conflict has been provided by recent work which points to the multiple technical flaws inherent in the pressure probe measurements (Steudle, 2001). Measurements conducted by Wei and colleagues (Wei et al., 2001; Wei et al., 1999a; Wei et al., 1999b) showed that the pressure probe works best for small volume measurements but its range of sensitivity was restricted to tensions less than 0.8 MPa. They enclosed the roots of plants in pressure chambers and measured the propagation of pressure pulses through the plant xylem. They observed that unless the tip was properly positioned, the pressure may drop to below zero but it does respond rapidly to pressure pulses. Hence an improperly positioned probe is unable to respond one a one-to-one basis to changes in pressure. However, it was possible to position the probe in such a way that its tip was properly situated in a functioning vessel, and under these conditions of proper hydraulic contact, any change in pressure on the plant showed a corresponding change in the microprobe measurement. Moreover, it was also possible to show that under these conditions, increase in transpiration due to increase incident light on the leaf caused a corresponding decrease in xylem pressure, as would be expected by the cohesion-tension theory of ascent of sap. Thus, the microprobe measurements offer not only a more accurate and non-invasive picture of the internal pressures in xylem, but they also point towards cohesion-tension as the likely mechanism for sap transport.

### Acknowledgments
We would like to thank Dr. Uday Pathre, National Botanical Research Institute for his helpful comments to this article.

Received 30 August 2011.

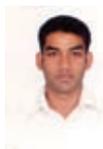

**Amit K. Singh** is currently as a Junior Research Fellow in the Insect flight laboratory at the National Centre for Biological Sciences, Bangalore. He received his Bachelor's degree in Biotechnology (2004–08) from Bangalore University. His areas of interest include plant movement and mechanics, organismal biomechanics, bio-optics and micro-fluidics.

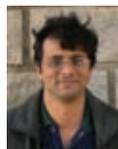

**Sanjay P. Sane** is an Assistant Professor at the National Centre for Biological Sciences, Bangalore since August, 2007. He received his Ph.D. from the Department of Integrative Biology at the University of California, Berkeley in 2001 followed by a post-doctorate at the University of Washington, Seattle. His primary area of interest is the physics and sensory neurobiology of insect flight. He is also interested in related aspects of comparative biomechanics, eco-physiology and evolutionary biology.